# Direct observation of photonic Landau levels and helical edge states in strained honeycomb lattices

Omar Jamadi[1*†], Elena Rozas[2], Grazia Salerno[3], Marijana Milićević[4], Tomoki Ozawa[5], Isabelle Sagnes[4], Aristide Lemaître[4], Luc Le Gratiet[4], Abdelmounaim Harouri[4], Iacopo Carusotto[6], Jacqueline Bloch[4] & Alberto Amo[1]

[1]*Université de Lille, CNRS, UMR 8523 – PhLAM – Physique des Lasers Atomes et Molécules, Lille, 59000, France*

[2]*Depto. de Física de Materiales e Instituto Nicolás Cabrera, Universidad Autónoma de Madrid (UAM), Madrid, 28049, Spain*

[3]*Center for Nonlinear Phenomena and Complex Systems, Université Libre de Bruxelles, CP 231, Campus Plaine, B-1050 Brussels, Belgium*

[4]*Université Paris-Saclay, CNRS, Centre de Nanosciences et de Nanotechnologies, 91120, Palaiseau, France*

[5]*Interdisciplinary Theoretical and Mathematical Sciences Program (iTHEMS), RIKEN, Wako, Saitama 351-0198, Japan*

[6]*INO-CNR BEC Center and Dipartimento di Fisica, Universita di Trento, 38123 Povo, Italy*

---

*Corresponding author.
†E-mail: omar.jamadi@univ-lille.fr

**Abstract**

We report the realization of a synthetic magnetic field for photons and polaritons in a honeycomb lattice of coupled semiconductor micropillars. A strong synthetic field is induced in both the $s$ and $p$ orbital bands by engineering a uniaxial hopping gradient in the lattice, giving rise to the formation of Landau levels at the Dirac points. We provide direct evidence of the sublattice symmetry breaking of the lowest-order Landau level wavefunction, a distinctive feature of synthetic magnetic fields. Our realization implements helical edge states in the gap between $n = 0$ and $n = \pm 1$ Landau levels, experimentally demonstrating a novel way of engineering propagating edge states in photonic lattices. In light of recent advances in the enhancement of polariton-polariton nonlinearities, the Landau levels reported here are promising for the study of the interplay between pseudomagnetism and interactions in a photonic system.

**Introduction**

One of the most remarkable consequences of the application of an external magnetic field to a two-dimensional electron gas is the quantum Hall effect. In this phenomenon, the parabolic band dispersion of the electron gas rearranges into discretized Landau levels[1]. Moreover, when the Fermi energy is located between two Landau levels, the system is insulating in the bulk and possesses conducting channels on the edge. These channels are so robust to disorder that they can be used for metrology applications[2].

The search for such effects in photonic systems has been a very active field in the past decade[3]. The implementation of Landau levels for photons is a solid-state inspired route to the study of artificial magnetic fields acting on photons. Thanks to the presence of photon nonlinearities in certain optical media, photonic lattices showing Landau levels would be particularly suited for the investigation of the interplay between pseudomagnetism and interactions[4–6]. In addition, the associated unidirectional edge channels are ideal to engineer photonic transport immune to backscattering in a chip[3,7]. The main difficulty in attaining this regime is that photons are largely insensitive to external magnetic fields.

Beyond the successful observation of chiral edge states for microwave, telecom and near-infrared wavelengths using magneto-optical materials[8–10], the realisation of topological bands for photons has so far relied on the engineering of synthetic magnetic fields. In this way, the Harper-Hofstadter model in coupled micro-resonators[11,12] and Landau levels in centimetre-size cavities[13] have been realised for photons without the need for any external magnetic field. A very straightforward approach to engineer synthetic magnetic fields was proposed by Guinea and co-workers when considering a layer of graphene[14,15]: by carefully engineering the strain applied to the honeycomb lattice, a gradient of hopping strength between adjacent sites is obtained, resulting in electrons experiencing a synthetic magnetic field. As time-reversal is not broken, the synthetic magnetic field has opposite signs at each of the two Dirac cones. Since this configuration requires only the local modification of the nearest-neighbour hopping energies across the honeycomb lattice[16,17], such strain-induced pseudomagnetic fields can be directly transferred to graphene analogues based on classical waves of different nature[18], from the photons/polaritons considered here to phonons. The resulting Landau-level structure is characterized by a zero-energy level ($n = 0$) and by a set of positive and negative quasi-flat levels ($n = \pm 1, \pm 2, \pm 3, ...$) with a square root energy spacing ($\epsilon_n \sim \pm\sqrt{n}$) inherited from the Dirac dispersion, similar to a graphene sheet exposed to a strong perpendicular magnetic field[19].

In photonics, this configuration was first implemented by Rechtsman et al. in a lattice of coupled waveguides subject to artificial trigonal strain[20]. The presence of flat Landau levels was inferred from the localisation dynamics of a point-like excitation at the edge of the lattice. However, the observation of the Landau-level spectra and the unique properties of the wavefunctions in the presence of a synthetic field, such as sublattice polarization[18,21] and helical propagation[22], have remained unexplored. More recently, the implementation of strain-induced Landau levels has been proposed[23,24] and reported[25] in macroscopic acoustic lattices.

In this article, we report direct evidence of photonic Landau levels and helical edge states emerging from a synthetic magnetic field engineered in strained honeycomb lattices for microcavity polaritons[26]. Unlike other photonic analogues of graphene, such as coupled waveguides[20,27], microwave resonators[28,29] and photorefractive crystals[30], polaritons allow direct access to the energy band structure and to the spatial distribution of the wavefunction. Taking full advantage of these properties, this system has been a versatile platform to bring to the photonics realm the single-particle physics of quasi-crystals[31] and graphene[10,32–34] and to engineer new types of Dirac cones[35] and other two-dimensional lattices[36,37]. Here, we follow the theoretical proposal of Salerno and co-workers[38] and use an artificial uniaxial strain to generate strong synthetic fields for exciton-polaritons. We observe Landau levels at the Dirac cone energy in the $s$ and $p_{x,y}$ orbital bands arising, respectively, from the coupling between the fundamental and first excited modes of each micropillar. For both sets of bands, we observe the localization of the $n = 0$ Landau level wavefunction in one sublattice, illustrating the specificity of pseudomagnetic fields compared to real ones (see Ref.[39] for an implementation in the microwave regime). We also report propagating helical edge states associated with the $n = 0$ Landau level, the synthetic field analogue of the chiral edge states responsible for transport in the quantum Hall effect. In light of the recent progress on polariton quantum effects[40,41], our platform provides an alternative route to twisted cavities[6,13] for the study of the fractional quantum Hall Laughlin states of strongly interacting photons anticipated in Ref.[42].

**Results**

To fabricate the strained polariton honeycomb lattices, we start from a microcavity grown by molecular beam epitaxy, composed of a $\lambda/2$ Ga$_{0.05}$Al$_{0.95}$As cavity embedded in two Bragg mirrors of 28 (top) and 40 (bottom) pairs of $\lambda/4$ alternating layers of Ga$_{0.05}$Al$_{0.95}$As/Ga$_{0.80}$Al$_{0.20}$As. Twelve 7-nm-wide quantum wells are positioned at the three central maxima of the electromagnetic field. At 10 K, the temperature of our experiments, the strong coupling regime between quantum well excitons and confined photons gives rise to polaritons, light-matter hybrid quasi-particles, characterised by a Rabi splitting of 15 meV. Though the experiments reported here are restricted to the low-power linear regime, the hybrid light-matter nature of polaritons has the potential to reach nonlinear regimes at high excitation densities.

The planar microcavity is processed by electron beam lithography and inductively coupled plasma etching down to the GaAs substrate to form honeycomb lattices of overlapping micropillars. Each micropillar has a diameter of 2.75 $\mu$m. Due to the additional lateral confinement provided by the refractive index contrast between the semiconductor and air, the micropillars act as artificial photonic atoms. The fundamental mode of their energy spectrum ($s$ mode) is cylindrically symmetric, similar to the $p_z$ orbitals in graphene. The first excited state is made of two antisymmetric modes, $p_x$ and $p_y$, with lobes oriented in orthogonal spatial directions, as sketched in Fig. 1(a). During the design of the lithographic mask, an overlap between micropillars is introduced to enable the hopping of polaritons between adjacent sites[43]. If the hopping from each pillar to its three nearest neighbours is equal and spatially independent ($t_1=t_2=t_3$, unstrained honeycomb lattice), the coupling of $s$ modes gives rise to two bands similar to the $\pi$ and $\pi^*$ bands of graphene (Fig. 1(b)), whereas the coupling of $p$ modes results in a set of four bands at higher energy (two of them are dispersive and intersect linearly, giving rise to Dirac cones).

To implement artificial strain in our lattices, we design the lithographic mask in such a way that the centre-to-centre distance between the micropillars along the links parallel to $x$ varies continuously between 1.9 and 2.7 $\mu$m, while the distance between the other links remains the same over the whole lattice (a=2.4 $\mu$m; $t_2 = t_3$, see Fig. 1(c)). This uniaxial gradient of interpillar distances is designed to yield a linear evolution of the hopping $t_1(x)$ in both the $s$ and $p$ bands (see Supplementary material):

$$t_1(x) = t\left(1 + \frac{x}{3a}\tau\right), \quad (1)$$

where $t = t_2 = t_3$ and $\tau$ quantifies the amount of strain. The strain induces a synthetic vector potential $\boldsymbol{A} = (A_x, A_y)$ with $eA_x = 0$ and $eA_y(x) = 2\xi\hbar\tau x/(9a^2)$, where $e$ is the electron charge and $\xi = \pm 1$ is the valley index associated with each of the two Dirac points[38] K and K'. The resulting artificial magnetic field $B_z = \partial_x A_y$ is directly proportional to the hopping gradient $\tau$ and has opposite signs in the two Dirac valleys K and K'. This is consistent with the fact that the strain-induced pseudomagnetic field does not break the time-reversal symmetry, which is the main difference from a magnetic field applied to graphene. We designed photonic honeycomb lattices with different hopping gradients resulting in artificial magnetic fields corresponding to values up to 3200 T for the $s$ bands and 1000 T for the $p$ bands when considering the electron charge and the lattice parameter of graphene. All the considered lattices share the same design: i) armchair termination for the bottom and top edges, ii) zigzag termination for the left and right edges, and iii) $t_1$ increases from the left to the right edge (i.e., positive gradient $\tau$). Figure 1(c) shows a scanning electron microscope image of the central region of one of the ribbons employed in the experiments.

Figure 1(d) shows the energy dispersion of the $s$ bands numerically calculated within a tight-binding model for a ribbon with $N_x = 41$ unit cells and a weak hopping gradient $\tau = 0.05$ along $x$. Along $y$, periodic boundary conditions have been assumed. In the simulation, the black colour indicates modes with a wavefunction localised in the bulk of the lattice, while other colours are used to indicate wavefunctions located either on the right or left edge. Photonic Landau levels $n$ appear in the vicinity of the Dirac cones $K$ and $K'$, revealing a square root energy structure typical of massless Dirac particles[19]: $\epsilon_n^s = E_0^s \pm t\sqrt{\tau|n|}$ with $E_0^s$ the energy of the Dirac points in the $s$ bands. The $n = 0$ Landau level appears at $E_0^s$, both at $K$ and $K'$. At higher and lower energies, higher-order ($|n| > 0$) Landau levels appear in the simulation. These levels present a small tilt caused by the spatial variation of the Dirac velocity along the horizontal direction[44,45]. Indeed, high-order Landau levels whose associated wavefunctions are centred in different positions of the lattice have slightly different energies[38].

The calculated bands in Fig. 1(d) also show the presence of edge states, depicted by the cyan and magenta colours. Between K and K', the coloured line linking the two $n = 0$ Landau levels corresponds to trivial edge states associated with the zigzag terminations. These states are also present in honeycomb lattices with homogeneous hopping[33]. Away from the K and K' points, the $n = 0$ Landau level evolves into two split bands with energy higher and lower than $E_0^s$. These bands correspond to propagating states located at the right edge of the lattice, as indicated by the magenta colour. Bearded terminations (not shown here) result in edge states located on the left edge[22]. Higher-order ($|n| > 0$) Landau levels give rise to propagating states located at the right and left edges. Since the time-reversal symmetry is preserved, edge states in different valleys propagate in opposite directions, resulting in helical transport. The goal of this article is to report on the experimental implementation of these photonic Landau levels and their associated helical edge states.

The experimental investigation of the lattices is realized by combining reciprocal- and real-space photoluminescence experiments using a non-resonant continuous wave laser at 740 nm. The laser is focused on an 8 $\mu$m diameter spot at the centre of the lattice by an aspherical lens ($N.A.= 0.5$). The low value of the pump power (0.2 mW) avoids any nonlinear effects. The real- and momentum-space resolved photoluminescence is measured using an imaging spectrometer coupled to a CCD camera. A polariser in front of the spectrometer selects the emission linearly polarized along the y axis.

First, we focus on the photonic Landau levels in the $s$ bands. Figure 2(a) displays the angle-resolved photoluminescence in the middle of the second Brillouin zone ($k_x = 4\pi/3a$) in a lattice without a hopping gradient ($\tau = 0$; $25 \times 18$ unit cells). This shows a band structure very similar to the $\pi$ and $\pi^*$ bands of graphene, with Dirac cones at their crossing. The dispersion is fitted (white line) by a tight-binding model considering nearest- and next-nearest-neighbour hopping ($t = 0.17$ meV, $t' = -0.08t$). The latter is an effective hopping that allows us to reproduce the asymmetry between $\pi$ and $\pi^*$ bands[32,33]. Physically, it originates from long-distance coupling mediated by the $p$ bands[46]. Figure 2(b) shows the measured dispersion of a lattice with artificial strain ($\tau = 0.56$, $B_z = 1400$ T) when summing the emission over all values of the transverse $k_x$ accessible with the collection lens for a given value of $k_y$ (parallel to the zigzag edge). The white lines indicate the theoretical dispersion of the discrete bands that originate from the splitting of the bulk bands in our finite lattice ($10 \times 18$ unit cells, 360 micropillars) with a high hopping gradient.

Figure 2(b) shows, by the white lines, the calculated dispersion using a tight-binding Hamiltonian that takes into account the nominal variation of $t_1(x)$ implemented in the lattice and an onsite energy offset $\Delta = 1.6t$ applied to the leftmost and rightmost columns of micropillars in the lattice. This onsite energy accounts for the additional confinement of the $s$ modes in the micropillars located at the edges, which only have two neighbouring pillars instead of three for those in the bulk, and therefore, their eigenfunction is more strongly confined than that in the bulk pillars. To simplify the model, we neglected any next-nearest-neighbour coupling. Close to $E_0^s$, at the Dirac points, the wavefunctions obtained from the model show spatial localisation at the centre of the lattice compatible with the $n = 0$ Landau level. In contrast, the bands extending away from the Dirac points around $E_0^s$ correspond to states localised at the zigzag edges. The full colour weighted version of the tight-binding fit, with information on the spatial position of the wavefunctions, is shown in the Supplementary material. Given the significant linewidth of the emission ($\sim 200$ μeV) comparable to the expected gap between the lowest-order Landau levels, the identification of the $n = 0$ level from momentum-space measurements is challenging.

To clearly identify the zeroth Landau level close to $E_0^s$, we perform real-space tomography of the wavefunctions at that energy. To do so, the surface of the lattice is imaged on the entrance slit of the spectrometer, and the photoluminescence is resolved in energy and real-space position $y$ for different positions $x$. The $x, y$ intensity distribution of the wavefunctions for any energy can then be reconstructed from the measured tensor $(x, y, E)$. Figure 2(c) shows the real-space emission at the energy of the Dirac points for the unstrained lattice ($\tau = 0$). The wavefunction presents a honeycomb pattern that extends over several lattice sites around the excitation spot (indicated by the white dashed line in the images) due to polariton propagation at the Dirac velocity. Figures 2(d) and (e) show the emitted intensity at $E_0^S$ for strained lattices with $\tau = 0.56$ ($B_z = 1400$ T) and $\tau = 1.26$ ($B_z = 3200$ T). In contrast to the unstrained case, in which the emission intensities from the A and B sublattices are equivalent, the presence of the pseudomagnetic field changes the intensity ratio between the two sublattices: for $\tau = 0.56$, a clear asymmetry between the A and B sublattices is visible in the bulk of the strained lattice, and for $\tau = 1.26$, the wavefunction is completely localised on the B sublattice. The measured and theoretical dispersions for $\tau = 1.26$ are shown in the Supplementary material.

Sublattice polarization is the main signature of the $n = 0$ Landau level under a pseudomagnetic field[18,38]. In the case of a real magnetic field applied perpendicularly to a graphene sheet, the field has the same sign at both Dirac points K and K'. The wavefunction of the $n = 0$ Landau level at the K point presents a non-zero amplitude in only one of the two sublattices, while at the K' point, the non-zero amplitude appears in the other sublattice. Overall, the $n = 0$ Landau level under a real magnetic field is a combination of both Dirac points and is distributed over both sublattices. In the case of a strain-induced pseudomagnetic field, the sign of the effective field at the K point is opposite to that at the K' point. Therefore, the wavefunction of the $n = 0$ Landau level is localised in the same sublattice for both Dirac points, and the specific sublattice A or B is determined by the gradient of the strain. For higher-order Landau levels ($n \neq 0$), the wavefunctions present a non-zero probability amplitude in both sublattices in both real and strain-induced fields. In the experiments, since an increase in the pseudomagnetic field leads to a larger gap between the $n = 0$ and higher-$n$ Landau levels, the strongest sublattice polarization is observed for the lattice with the highest hopping gradients $\tau$ (Fig. 2(d) and (e)). This sublattice polarization provides unambiguous identification of the zeroth Landau level.

The appearance of the zeroth Landau level is also visible in the measured density of states shown in Fig. 2(f) for different hopping gradients. The density of states is obtained from the integration in space of the emitted light at a given energy. The $n = 0$ Landau level manifests as a peak of high density of states at $E_0^s$, as a consequence of the associated degenerate flat band, when the gradient $\tau$ increases. To avoid any possible contribution of the trivial zigzag edge states also present at $E_0^s$, in Fig. 2(f), we have excluded from the integration the emission of the two leftmost and two rightmost columns of micropillars in lattices with $\tau = 0.56$, $\tau = 0.96$ and $\tau = 1.26$.

One should also expect the presence of propagating edge states emerging from the pseudomagnetic field at energies between the different Landau levels (see Fig. 1(d)). However, given the emission linewidth, propagating edge states mix with bulk states and cannot be evidenced in the $s$ bands. For this reason, we turn our attention to the $p$ bands, which are also present in the fabricated lattices. As the nearest-neighbour coupling $t_L$ for the $p$ orbitals oriented along the link between adjacent micropillars is four to five times larger than that for the $s$ bands[32], we expect a larger Landau-level spacing and the possibility of identifying helical edge states emerging from the artificial magnetic field.

Figure 3(a) displays the angle-resolved photoluminescence of the $p$ bands when exciting an unstrained lattice ($\tau = 0$, $B_z = 0$ T) at its centre. The dispersion is recorded as a function of $k_y$ for a value of $k_x =$ $4\pi/3a$, passing through the centre of the second Brillouin zone. One can observe a flat band at low energy and two intermediate dispersive bands with Dirac crossings at the K and K' points. By implementing the hopping gradient of Eq. (1) in the longitudinal hopping $t_{L,1}(x)$ for $p_x$ orbitals parallel to the horizontal links, we expect to engineer an artificial magnetic field around the Dirac points similar to that of the s bands: photonic Landau levels should appear with an energy spectrum $\epsilon_n^p = E_0^p \pm (t_L/2)\sqrt{\tau|n|}$ (see Supplementary material), where $E_0^p$ is the energy of the Dirac points in the $p$ bands. Figure 3(b) shows the photoluminescence for the strained lattice shown in Figs. 2(b) and (d) obtained when summing the emission over all accessible values of the transverse $k_x$ for each value of $k_y$. For the $p$ bands, the designed centre-to-centre separation results in a hopping gradient of $\tau = 0.2$ ($B_z = 500$ T). The flat bands around zero energy that are visible for $k_y \in [-2, -4]k_{y0} \cup [2,4]k_{y0}$ delimited by the Dirac points at $k_y = \pm 2k_{y0}$ and $\pm 4k_{y0}$ (with $k_{y0} = 2\pi/3\sqrt{3}$) correspond to the trivial zigzag edge states[34], similar to those shown for the s bands in Fig. 1(d). At the position of the Dirac points, two energies $\epsilon_{0-}^p = E_0^p - 0.07$ meV and $\epsilon_{0+}^p = E_0^p + 0.07$ meV can be defined for the $n = 0$ Landau level (see the zoomed area in Fig. 3(c)). This apparent splitting of the Landau level originates from the finite size of the lattice; it is inherent to any experimental realization and should decrease for larger lattices. Interestingly, the $n = -1$ Landau level is also visible in Fig. 3(c). The nearest-neighbour tight-binding simulations shown in Fig. 3(b) and 3(c) as solid lines qualitatively reproduce the observed dispersion for $t_L = -0.85$ meV and assuming that the hopping $t_T$ between the $p$ orbitals oriented perpendicular to the link is zero[32]. Due to the naturally stronger confinement of the $p$ modes in the pillars, in this case, we do not assume any onsite energy offset at the edge micropillars.

To confirm the identification of the $n = 0$ Landau level, we compare the wavefunction at the Dirac energy of an unstrained lattice ($B_z = 0$ T, Fig. 3(e)) with the wavefunctions recorded at $E = \epsilon_{0-}^p$ for $\tau = 0.2$ ($B_z = $ 500 T, Fig. 3(f)). For zero pseudomagnetic field, the wavefunction is delocalized over several lattice sites and presents a honeycomb pattern made of $p_{x,y}$ orbitals. Analogous to that of the $s$ bands, the intensity distribution in the bulk of the strained lattice presents a B sublattice polarization, a clear signature of the $n = 0$ Landau level emerging from a pseudomagnetic field. One can notice a stronger contribution of $p_x$ over $p_y$ orbitals on the left side of the $n = 0$ Landau level wavefunctions. This feature arises from the larger onsite energy of $p_x$ orbitals inherited from the stronger confinement of the $p$ orbitals along the direction of the hopping gradient when $t_{L,1}(x) < t_L$. If the hopping gradient is further increased, the sublattice polarization becomes more evident, as displayed in Fig. 3(g)) for $\tau = 0.4$.

The measured density of states shown in Fig. 3(d) is consistent with the emergence of the degenerate $n = 0$ Landau level when the strain is increased. The wavefunction of the $n = -1$ Landau level visible in Fig. 3(c) is shown in Fig. 3(h). As expected, the intensity on sublattices A and B is similar in this case since the sublattice polarization is exclusive to the zeroth Landau level.

The larger hopping strength between the $p$ orbitals than that between the $s$ bands opens up the possibility of observing propagating edge states in the gap between the $n = 0$ and $n = \pm 1$ Landau levels, as shown in the tight-binding simulations in Fig. 1(d) for the $s$ bands. Figure 3(c) highlights the presence of two such propagating edge states close to the K point in the gaps between $p$ Landau level $n = 0$ and Landau levels $n = \pm 1$. The tight-binding simulations depicted in the lower panel of Fig. 3(c) and in the Supplementary material show the continuous evolution of the $n = 0$ Landau-level bands spatially located in the bulk of the lattice (black colour) into the propagating edge states spatially located on the edge (magenta, right edge), as expected from a two-dimensional system subject to an artificial magnetic field. Similar edge states with opposite group velocities are also visible at the other Dirac cone.

The propagation of these edge states is experimentally studied by placing the excitation spot on the zigzag edges of a strained lattice with a hopping gradient $\tau = 0.28$ ($B_z = 700$ T). The measured dispersions, centred and zoomed on the zeroth Landau level, are shown in Figs. 4(a) and (b) for the excitation on the left and right edges, respectively. The dispersions exhibit very different features depending on the excited edge. Figure 4(a) displays the dispersion corresponding to the excitation of the left zigzag edge, and it is dominated by a flat band at $E_0^p$ associated with non-propagating zigzag edge states[34]. These modes and the $n = -1$ Landau levels at lower energy are separated by a gap (a similar gap is also visible just above $E_0^p$). In contrast, Fig. 4(b), which corresponds to the excitation of the right zigzag edge, is dominated by dispersive modes near the Dirac points $k_y = \pm 2k_{y0}$ with opposite group velocities, lying exactly in the gap between the zeroth and $n = -1$ Landau levels.

These dispersive modes correspond to the propagating edge states linked to the zeroth Landau level depicted in Fig. 1(d) for the $s$ bands. Since they have opposite group velocities in each valley, they propagate in opposite directions. This helical propagation is revealed in Fig. 4(d) when measuring the emitted intensity for an energy lying between $\epsilon_{0-}^p$ and $\epsilon_{-1}^p$ (horizontal red line in Figs. 4(a) and (b)). At the right edge, polaritons flow from the excitation spot towards the top and bottom corners of the lattice. When the left edge is excited, as shown in Fig. 4(c), the measured intensity shows a fast exponential decay outside the excitation spot, evidencing the absence of propagating edge states. This left/right propagation asymmetry is well reproduced by driven dissipative calculations (see Supplementary material). The intensity profiles measured along the left and right zigzag edges (yellow rectangles in (c) and (d)) are reported in Fig. 4(e). The extracted propagation lengths are $L_l = 5$ $\mu$m and $L_r = 17$ $\mu$m for the left and right edges, respectively. Based on the propagation length $L_r$ of the right edge state and the group velocity $v_g = 2.4$ $\mu$m.s$^{-1}$ measured in Fig. 4(b), we find a lifetime of $\tau \approx 7$ ps for polaritons in the propagating state, in agreement with previous experiments on polariton honeycomb lattices etched from the same wafer[35].

To understand why only the right zigzag edge supports the propagation of the zeroth Landau level, one has to consider the chiral symmetry of our system and the peculiar form of the zeroth Landau level wavefunction emerging from the pseudomagnetic field[22]. As the zeroth Landau level wavefunction is entirely localised on the B sublattice, its energy is pinned to $E_0^p$ (the onsite energy of the isolated $p$ orbitals) by the chiral symmetry. Consequently, the only way for the zeroth Landau level to evolve into a dispersive state with $E \neq E_0^p$ is to combine with other zero-energy edge states localised on the A sublattice. In our case, the role of these zero-energy states is played by trivial edge states associated with the zigzag termination of the honeycomb lattice. These trivial edge states are localised on the A sublattice at the right edge and on the B sublattice at the left edge. For this reason, propagating edge states associated with the $n = 0$ Landau level appear only at the right edge for the sign of the hopping gradient that we consider. Note that by choosing other terminations, it is possible to engineer propagating edge states only at the left edge, propagating edge states at both edges simultaneously, or no propagating edge states at all[22,47]. This behaviour is a consequence of the sublattice symmetry breaking of the $n = 0$ Landau level emerging from the presence of the pseudomagnetic field and is very different from the unidirectional chiral edge states that appear in graphene under an external magnetic field, with a propagation direction independent of the type of edge and robust to local disorder.

**Discussion**

In this work, we have reported a direct measurement of the $n = 0$ Landau level wavefunction distributions in a photonic honeycomb lattice subject to a synthetic magnetic field. We have provided unprecedented evidence of the dispersion of helical edge states induced by an artificial magnetic field and their emergence from the fundamental Landau level at the edges of the sample. In combination with recent advances in the enhancement of polariton-polariton interactions[40,41,48], our realisation is promising for the prospect of studying strongly correlated photonic phases in systems with a higher degeneracy than those in recent reports for microwave[49] and twisted cavities[6]. The helical edge states we have unveiled provide a new playground to design topological photonic channels in a chip without any external magnetic field. They present a high degree of versatility, as their existence is controlled by the termination of the lattice and, similar to the interface states emerging from the valley Hall effect[50], they should in principle be free of backscattering in the presence of any disorder that does not mix the two non-equivalent Dirac cones.

**Methods**

The different hopping gradients $\tau$ are designed during the realization of the lithographic mask by accurately tuning the distance $d$ between the centres of the pillars forming a horizontal link. The distance between the centres of the pillars forming angled links with respect to the horizontal direction remains constant ($d = 2.4 \mu$m). To determine the dependence of the hopping on the centre-to-centre distance $d$, we solve the time-independent Schrödinger equation in two dimensions, accounting for the in-plane shape of the micropillars for different centre-to-centre distances[51] and assuming infinite potential out of the micropillars. We take $D = 2.75$ $\mu$m as the diameter of the two micropillars, the same as that engineered in the considered strained lattices. From the spectrum of eigenmodes, we extract the bonding-antibonding splitting for both the $s$ and $p$ modes as a function of the centre-to-centre distance. We assume that the splitting for each set of modes corresponds to twice the hopping amplitudes $t_s$ and $t_p$ in the tight-binding model. We perform a linear fit of $t_s$ as a function of $d$ and a parabolic fit for $t_p$, resulting in the following dependences:

$$d = -1.3 \times t_s + 2.73, \quad for\ s\ bands,$$

$$d = -0.75 \times t_p^2 + 0.34 \times t_p + 2.74, \quad for\ p\ bands.$$

We use the above equations to design the centre-to-centre distances along the horizontal direction in the considered lattices. Each lattice is designed to have a strictly linear gradient $\tau$ either in the $s$ or in the $p$ bands. However, even if a lattice is designed to have a linear hopping gradient in one of the two types of bands, the hopping gradient in the other type of band is also linear to a very good approximation. Supplementary Fig. S1 shows the exact centre-to-centre distances along the horizontal direction employed in the samples used in our study (black dots) along with the $s$ and $p$ band hopping in the horizontal links expected from the above equations. Lattices (a), (c) and (d) were designed with a linear gradient in the $p$ bands, while (b) was designed with a linear gradient in the $s$ bands. Nevertheless, the expected hopping gradient for the other type of band is also linear to a very good approximation (the figure shows the $R^2$ errors to linear fits). Therefore, our strained lattices implement homogeneous synthetic magnetic fields for both the $s$ and $p$ bands at the Dirac points.

**Acknowledgements**

This work was supported by the ERC grant Honeypol, the H2020-FETFLAG project PhoQus (820392), the QUANTERA project Interpol (ANR-QUAN-0003-05), the French National Research Agency project Quantum Fluids of Light (ANR-16-CE30-0021), the French government through the Programme Investissement d'Avenir (I-SITE ULNE / ANR-16-IDEX-0004 ULNE) managed by the Agence Nationale de la Recherche, the French RENATECH network, the Labex CEMPI (ANR-11-LABX-0007), the CPER Photonics for Society P4S and the Métropole Européenne de Lille (MEL) via the project TFlight. E.R. acknowledges financial support from FPI Scholarship No. BES-2015-074708. G.S. is supported by funding from the ERC Starting Grant TopoCold. T.O. is supported by JSPS KAKENHI Grant Number JP18H05857, JST PRESTO Grant Number JPMJPR19L2, JST CREST Grant Number JPMJCR19T1, the RIKEN Incentive Research Project, and the Interdisciplinary Theoretical and Mathematical Sciences Program (iTHEMS) at RIKEN.

**Conflict of interest**

The authors declare that they have no conflicts of interest.

**Contributions**

O. J. and E. R. performed the experiments and analysed the data; G. S., T. O. and I. C. performed the simulations and developed the theoretical model; M. M. designed the sample; I. S., A. L., L. L. G., and A. H. fabricated the samples; O. J. wrote the manuscript with assistance from all the co-authors; and I. C., J. B. and A. A. supervised the work.

227402 (2018).
49  Ma, R. C. *et al*. A dissipatively stabilized Mott insulator of photons. *Nature* **566**, 51-57 (2019).
50  Noh, J. *et al*. Observation of photonic topological valley hall edge states. *Physical Review Letters* **120**, 063902 (2018).
51  De Vasconcellos, S. M. *et al*. Spatial, spectral, and polarization properties of coupled micropillar cavities. *Applied Physics Letters* **99**, 101103 (2011).

**Figures**

Figure 1

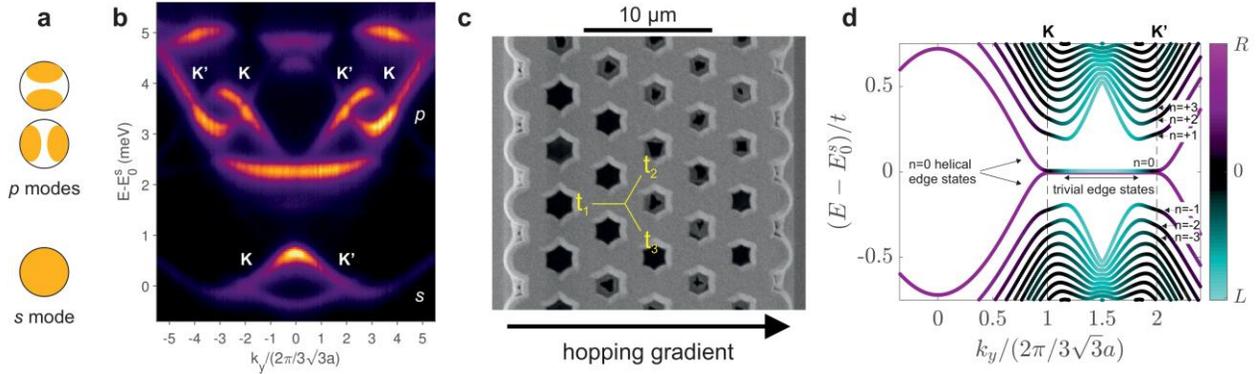

**Figure 1. Strained honeycomb lattice. a** Sketch of the real-space distribution of the $s$ and $p$ modes in a single micropillar. **b** Spectrally resolved far-field emission of an unstrained lattice showing the $s$ and $p$ bands for $k_x = 4\pi/3a$. $E_0^s = 1565.25$ meV is the energy of the Dirac points in the $s$ bands. **c** Scanning electron microscope image of a strained honeycomb lattice with a gradient along the $x$ direction: $\tau = 0.96$ for the $s$ bands and $\tau = 0.28$ for the $p$ bands. **d** Tight-binding calculation of the band structure of a uniaxially strained ribbon containing $N_x = 41$ unit cells and a gradient $\tau = 0.05$ along $x$ and zigzag terminations. Periodic boundary conditions are assumed along $y$. Landau levels appear around the Dirac points K and K' for $k_y = \pm 2\pi/3\sqrt{3}a$. The $n = 0$ Landau level is degenerate with non-propagating trivial edge states on both zigzag edges, while propagating edge states appear only on the right zigzag edge. The colours are defined by the mean position of the wavefunction of each state: magenta for states localised on the right edge, cyan for states localised on the left edge, and black for bulk states.

Figure 2

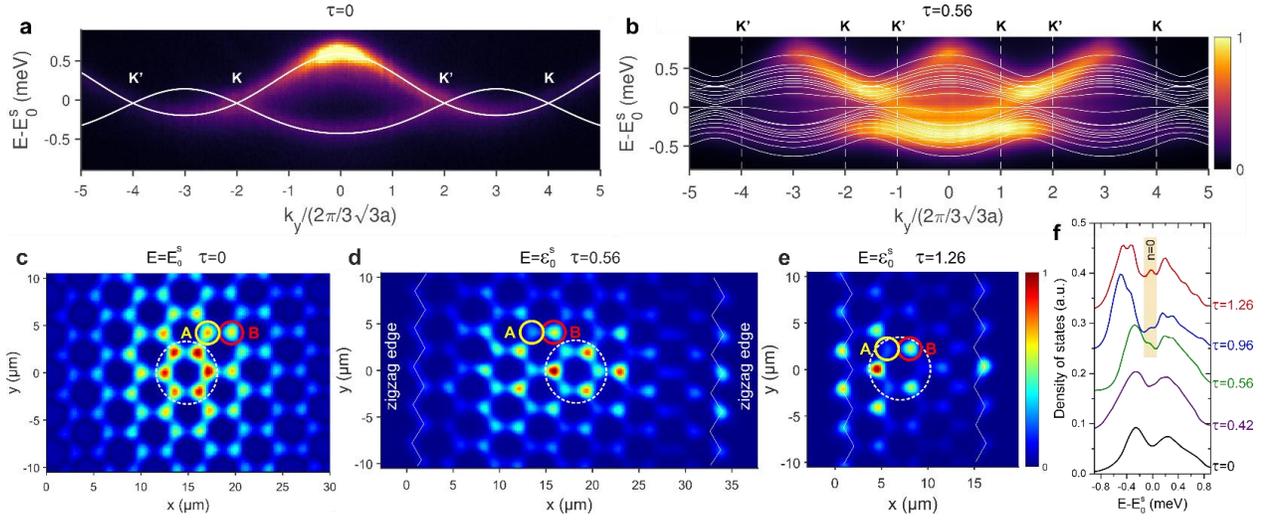

**Figure 2. Photonic Landau levels in *s* bands. a** Spectrally resolved far-field emission of the *s* bands from an unstrained lattice for $k_x = 4\pi/3a$. **b** Spectrally resolved far-field emission summed over all values of $k_x$ accessible in our setup for a strained lattice with a gradient $\tau = 0.56$. The corresponding pseudomagnetic field is $B_z = 1400$ T. The unshifted positions of the Dirac cones K and K' are indicated by vertical dashed lines. The Dirac energies are located at $E_0^s = 1565.25$ meV for **a** and $E_0^s = 1567.7$ meV for **b**. For each case, the white solid lines are theoretical fits using the tight-binding Hamiltonian with: $t = 0.17$ meV and $t' = -0.08t$ for **a**; $t = 0.17$ meV, $t' = 0$ meV, a hopping gradient $\tau = 0.56$, and an onsite energy of $+1.6t$ in the leftmost and rightmost pillar columns to account for the additional confinement of edge micropillars for **b**. **c-e** Measured real-space photoluminescence intensity at the energy of the Dirac cone for hopping gradients $\tau = 0$ **c**, $\tau = 0.56$ **d**, and $\tau = 1.26$ **e**. The position of the excitation spot is marked by white dashed lines. The colour scale of each panel has been independently normalised to its maximum value. **f** Measured density of states for different hopping gradients. For the strained lattices with $\tau = 0.56$, $\tau = 0.96$ and $\tau = 1.26$, the emission of the two leftmost and rightmost columns of micropillars has been excluded to suppress the contribution of the zigzag edge states. Each curve is vertically offset by 0.1 for clarity.

Figure 3

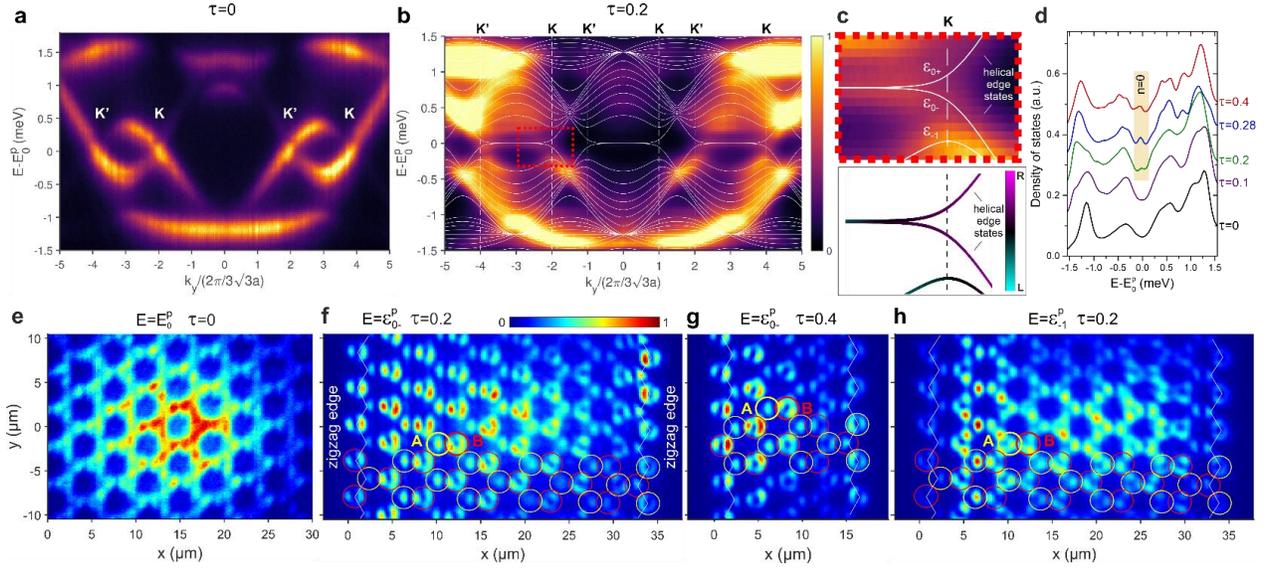

**Figure 3. Photonic Landau levels in $p$ bands. a** Spectrally resolved far-field emission of the $p$ bands from an unstrained lattice for $k_x = 4\pi/3a$. **b** Spectrally resolved far-field emission summed over all values of $k_x$ accessible in our setup for a strained lattice with a gradient $\tau = 0.2$. The corresponding pseudomagnetic field is $B_z = 500$ T. The Dirac energies are located at $E_0^p = 1568.5$ meV for **a** and $E_0^p = 1570.7$ meV for **b**. The positions of the Dirac cones K and K' are indicated by vertical dashed lines. **c** Top panel: zoom on the Landau levels $n = 0$ and $n = -1$. The contrast is increased to improve the visibility. The white lines in **b,c** are theoretical fits using the tight-binding Hamiltonian with $t_L = -0.85$ meV and $t_T = 0$ meV. Bottom panel: coloured version of the theoretical fit used in the top panel. The colour of each state is defined by the mean position of its wavefunction. **d** Measured density of states for different hopping gradients. For the strained lattices with $\tau = 0.2$, $\tau = 0.28$ and $\tau = 0.4$, the emission of the two leftmost and rightmost columns of micropillars has been excluded to suppress the contribution of the zigzag edge states. Each curve is vertically offset by 0.08 for clarity. **e-g** Measured real-space photoluminescence intensity for hopping gradients $\tau = 0$ **e**, $\tau = 0.2$ **f**, and $\tau = 0.4$ **h** at the energy of the Dirac cone. These lattices are, respectively, those shown in Fig. 2(c)-(e). The positions of the excitation spot are identical to those in Fig. 2. The $n = 0$ Landau level wavefunctions in **f** and **g** are characterized by a B sublattice polarization in the bulk. For clarity, circles show the positions of some of the pillars in the lattice. **h** Measured real-space photoluminescence intensity of the $n = -1$ Landau level shown in **c**. The colour scale of each panel has been independently normalised to its maximum value.

Figure 4

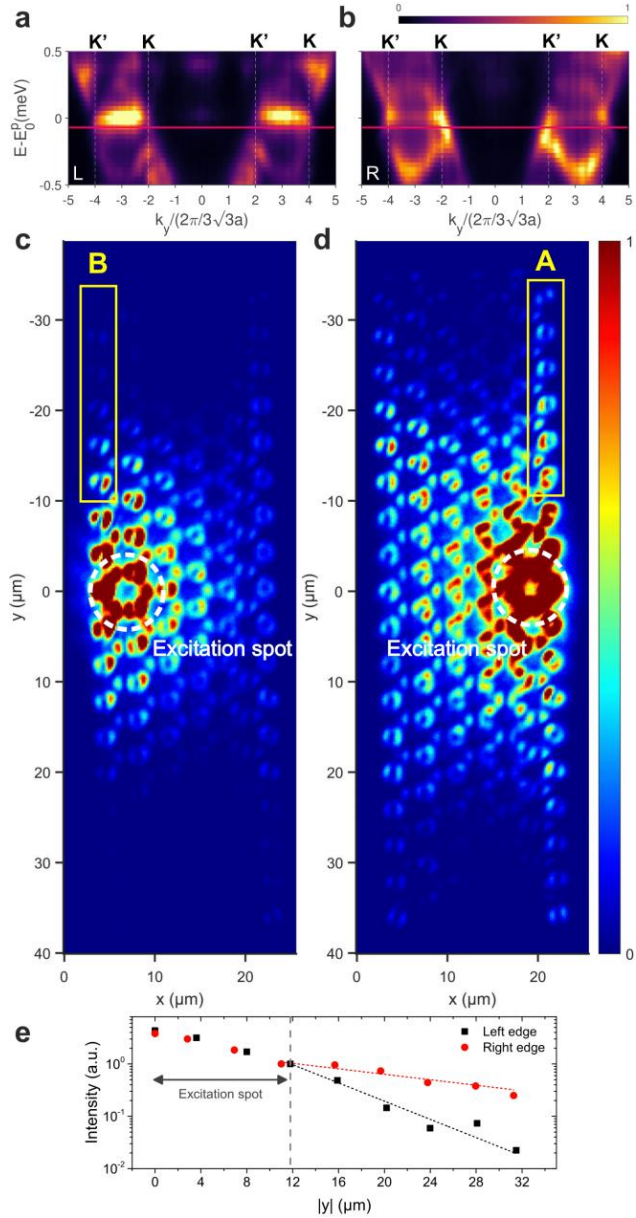

**Figure 4. Zeroth Landau level propagating edge states. a-b** Spectrally resolved far-field emission centred and zoomed on the zeroth Landau level ($p$ bands) of a strained lattice with a gradient $\tau = 0.28$ ($B_z = 700$ T) for a left **a** and right **b** edge excitation. $E_0^p = 1571.3$ meV. The positions of the Dirac cones K and K' are indicated by vertical dashed lines. **c-d** Measured real-space photoluminescence intensity pattern for an excitation localized on the left edge **c** and on the right edge **d**. The emission at an energy lying in the gap between $\epsilon_{0-}^p$ and $\epsilon_{-1}^p$ (horizontal red line in **a** and **b**) is selected. The position of the excitation spot is marked by white dashed lines. **e** Maximum intensity measured on each pillar above the excitation spot (yellow rectangles in **c** and **d**) for left and right edge excitation.